\documentclass[final,5p
,twocolumn,authoryear]{elsarticle}

\usepackage{amsmath,amsfonts,amssymb,mathtools}
\usepackage{hyperref}
\usepackage[nameinlink,capitalise]{cleveref}
\usepackage{bm,slashed}
\usepackage{xcolor}

\usepackage[scr]{rsfso}

\newcommand{\pd}{\partial}
\newcommand{\de}{\mathrm{d}}
\newcommand{\cH}{\mathcal{H}}

\renewcommand{\pd}{\partial}

\newcommand{\bmag}{\mathcal{Q}}
\newcommand{\bevo}{\mathcal{E}}
\renewcommand{\vec}[1]{\bm{#1}}
\newcommand{\tens}[1]{{\sf #1}}
\newcommand{\lmin}{\ell_{\rm min}}
\newcommand{\lmax}{\ell_{\rm max}}
\newcommand{\fsky}{f_{\rm sky}}
\DeclareMathOperator{\tr}{tr}

\usepackage{cancel}

\def\review#1{\textcolor{black}{#1}}

\let\Gamma\varGamma
\let\Delta\varDelta
\let\Theta\varTheta
\let\Lambda\varLambda
\let\Xi\varXi
\let\Pi\varPi
\let\Upsilon\varUpsilon
\let\Phi\varPhi
\let\Omega\varOmega

\journal{Physics of the Dark Universe}

\begin{document}

\begin{frontmatter}



\title{Detecting Local and Integrated Relativistic Effects by Multi-Tracing a Single Galaxy Population in Harmonic Space} 


\author[label1,label2]{Marco Novara}
\ead{marco.novara@edu.unito.it}
\author[label1,label2]{Federico Montano}
\ead{federico.montano@unito.it}
\author[label1,label2,label3,label4]{Stefano Camera}
\ead{stefano.camera@unito.it}

\affiliation[label1]{organization={Dipartimento di Fisica, Università degli Studi di Torino},
             addressline={Via P.\ Giuria 1},
             city={Torino},
             postcode={10125},
             country={Italy}}
\affiliation[label2]{organization={INFN -- Istituto Nazionale di Fisica Nucleare, Sezione di Torino},
             addressline={Via P.\ Giuria 1},
             city={Torino},
             postcode={10125},
             country={Italy}}
\affiliation[label3]{organization={INAF -- Istituto Nazionale di Astrofisica, Osservatorio Astrofisico di Torino},
             addressline={Strada Osservatorio 20},
             city={Pino Torinese},
             postcode={10025},
             country={Italy}}
\affiliation[label4]{organization={Department of Physics \& Astronomy, University of the Western Cape},
             city={Cape Town},
             postcode={7535},
             country={South Africa}}

\begin{abstract}
Measuring relativistic effects on cosmological scales would provide further confirmation of the validity of general relativity in the still poorly tested condition of weak gravity. 
Despite their relevance, relativistic imprints in the distribution of galaxies on large scales have so far eluded detection, mainly because they are stronger on the largest cosmic scales, which are plagued by cosmic variance.
Expanding on previous works, we study galaxy clustering by subdividing a galaxy population into two sub-samples---bright and faint---and we here focus on their two-point correlation function in harmonic space, i.e.\ via the angular power spectrum. Thanks to such a split in magnitude and by exploiting the multi-tracer technique, we are able to boost the impact of the relativistic contributions.
We first focus on the leading relativistic contribution given by the Doppler effect and show that, with a carefully tailored luminosity cut, it can be detected. 
Then, we look at the sub-dominant effects predicted by general relativity and quantify how their statistical significance, as yet undetectable, varies with redshift binning and survey specifications.
As case studies, we consider in our forecasts a bright galaxy sample at low redshift, an H$\alpha$ emission-line galaxy survey at intermediate redshifts, and high-redshift Lyman-break galaxies at high redshift.
\end{abstract}



\begin{keyword}
Galaxy Clustering \sep LSS \sep General Relativity \sep Angular power spectrum \sep Relativistic Doppler \sep Relativistic effects \sep Multi-tracer


\end{keyword}

\end{frontmatter}




\section{Introduction} \label{sec1intro}
Cosmological inference has found an invaluable mine of information in the large-scale structure of the universe. Current and future galaxy surveys are going to measure the distribution of biased tracers with unprecedented precision, allowing us to test the finest predictions of our theoretical models \citep{2024arXiv240513491E,2022AJ....164..207D,2024arXiv240305398M,2022arXiv220904322S}. Armed with the upcoming data, sampled on huge volumes, we will be able to tackle large-scale effects, such as local Primordial non-Gaussianity and relativistic contributions. 

General Relativity (GR) has been established as the standard description of gravitational interaction and offers the foundations for the Lambda Cold Dark Matter ($\Lambda$CDM) concordance model. Besides $\Lambda$CDM being challenged by several issues, GR itself is yet to be probed on cosmological scales. The weak gravitational field regime determinates galaxy clustering's statistical properties via a plethora of effects that act on the largest scales. These effects arise from galaxy surveys using redshift measurements to infer distances, and are expected to be detected as soon as we achieve enough statistical sampling on the very scales affected by cosmic variance.  

Overcoming \citet{10.1093/mnras/227.1.1}'s Newtonian description of the redshift-space distortion (RSD), the firsts theoretical derivations---within GR---of the observed galaxy overdensity was provided by \citet{2010PhRvD..82h3508Y,2011PhRvD..84f3505B,2011PhRvD..84d3516C}. They showed that, on top of the Kaiser RSD corrections, our theory of gravity predicts several other local and integrated along-the-line-of-sight effects---usually referred to as projection effects---accounting for observations being carried out on our past light cone. Among those, only lensing magnification has been measured so far \citep[see e.g.\ ][]{1998MNRAS.294..291M,1997ApJ...477...27B,2003ApJ...589...82G,2005MNRAS.359..741M,2005ApJ...633..589S,2023MNRAS.523.3649E}, therefore there is a compelling interest in today's cosmological studies in looking at them as possible confirmations of GR on cosmological regimes. 
However, new strategies need to be developed to detect the relativistic contributions, because they are sub-dominant compared to the Newtonian terms and relevant only on the large scales, where the lack of observable modes (i.e.\ cosmic variance) limits containing power. 

In this article, we argue that a measurement of a relativistic signature could be achieved in the future by analysing the harmonic space power spectrum of two sub-samples, a faint and a bright one, of the same galaxy population.
We estimate the statistical significance of the GR contributions at both leading-order---that is, the relativistic Doppler effect---and second-order---sourced by local and integrated metric potentials, for various actual and planned galaxy surveys, thus ranging from $z\approx0.1$ to $z\approx5$. Building on \citet{2023arXiv230912400M,2024PDU....4601634M}, we apply a multi-tracer approach, splitting a galaxy population into two sub-samples according to luminosity. Such an approach combines information coming from different biased tracers that sample the same underlying (dark) matter distribution \citep{2009PhRvL.102b1302S,2009JCAP...10..007M,2013MNRAS.432..318A,2022JCAP...04..013A}, therefore allowing us to mitigate cosmic variance. Previous researches have outlined the capability and formalism of multi-tracer analyses in harmonic space \citep{2014MNRAS.442.2511F,2015ApJ...812L..22F,2015PhRvD..92f3525A,2021JCAP...12..003F,2021JCAP...11..010V,2021MNRAS.502.2952T,2022JCAP...08..073A}. 
For the first time, we propose to use the luminosity cut technique in angular power spectrum analysis, while already existing works rely either on configuration \citep{Bonvin_2014,2014PhRvD..89h3535B,2016JCAP...08..021B,2017JCAP...01..032G,2023arXiv230604213B,2023MNRAS.519L..39S} or Fourier space \citep{2023arXiv230912400M,2024PDU....4601634M,2025PhRvD.111l3559C,Paper2Sam}. We include in our harmonic-space power spectra all of the projection effects, so that our model will describe the actual observations. Unlike Fourier space power spectrum and galaxy two-point correlation function measurements, wide-angle effects \citep{2018MNRAS.476.4403C} are, by construction, accounted for in harmonic space and do not represent much of a concern. 

Furthermore, given the intrinsic redshift and sample dependence of the contributions to the power spectra, we focus on three different galaxy populations. We adopt as benchmarks: a DESI-like bright galaxy sample (BGS) at low-$z$ ($0.05\leq z \leq 0.55$) \citep{2022AJ....164..207D}; an H$\alpha$ emission-line galaxy (ELG) sample at intermediate redshift ($0.9\leq z \leq 1.8$), akin to the spectroscopic sample targeted by the \textit{Euclid} satellite \citep{2024arXiv240513491E} or the \textit{Nancy Grace Roman} space telescope \citep{2019arXiv190205569A}; and a MegaMapper-like Lyman-break galaxy (LBG) target at $2.0 \le z \le 5.0$ \citep{2022arXiv220904322S}. This allows us to provide, for the first time, consistent forecasts over the extensive redshift interval $z\lesssim 5.0$, coherently accounting for the different behaviour of each target, without relying on simplified, general assumptions as has been commonly done in the literature.

We organise this paper as follows. In \cref{sec:fluctuation}, we review the relativistic expression of the linear galaxy number counts and how this translates into the observed angular power spectrum. In \cref{sec:MT}, we introduce the details of our multi-tracer analysis, specifying the galaxy populations considered. Our forecasts are presented in \cref{sec:results}, and our conclusions are given in \cref{sec:conclusions}.

\section{Linear fluctuations in galaxy number counts} \label{sec:fluctuation}
At linear order in cosmological perturbations, the density contrast of galaxy number counts, \(\Delta\), contains a number of terms due to the apparent galaxy positions being inferred from the observed redshift---which, in turn, is perturbed with respect to its value in a homogeneous and isotropic universe. Some of these terms are well known, first and foremost RSD \citep[see e.g.]{10.1093/mnras/227.1.1}, which induces an anisotropy along the line-of-sight direction. Acting orthogonally to it is instead the effect of lensing magnification, the more important the deeper the survey \citep{2021JCAP...04..055J}. Then, there are a score of other terms, collectively referred to as relativistic effects \citep{2009PhRvD..80h3514Y,2010PhRvD..82h3508Y,2011PhRvD..84d3516C,2011PhRvD..84f3505B}.

Quantitatively, we can write fluctuations in galaxy number counts as a function of their comoving position \(\vec x=(x_\|,\vec x_\perp)\), where \(x_\|\) is aligned with the line of sight, as
\begin{alignat}{2}
    \Delta(\vec x) 
    &= b\,\delta(\vec x) - \pd^2_\|\Theta(\vec x)/\cH
    &&\qquad \makebox[2cm][r]{\text{(density + RSD)}} \nonumber \\[4pt]
    &\quad - \alpha\,\pd_\|\Theta(\vec x) + (\bevo-3)\,\cH\,\Theta(\vec x)
    &&\qquad \makebox[2cm][r]{\text{(Doppler)}} \nonumber \\[4pt]
    &\quad + 2\,(1-\bmag)\,\kappa(\vec x)
    &&\qquad \makebox[2cm][r]{\text{(lensing)}} \nonumber \\[4pt]
    &\quad + \Phi'(\vec x)/\cH
    &&\qquad \makebox[2cm][r]{\text{(Sachs-Wolfe)}} \nonumber \\[4pt]
    &\quad + (1+\alpha)\,\varPsi(\vec x) - 2\,(1-\bmag)\,\Phi(\vec x)
    &&\qquad \makebox[2cm][r]{\text{(potentials)}} \nonumber \\[4pt]
    &\quad + \frac{2\,(1-\bmag)}{x_\|}\,\int_0^{x_\|}\de r\,\Upsilon(r,\vec x_\perp)
    &&\qquad \makebox[2cm][r]{\text{(time delay)}} \nonumber \\[4pt]
    &\quad + 2\,\alpha\,\int_0^{x_\|}\de r\,\Upsilon'(r,\vec x_\perp)
    &&\qquad \makebox[2cm][r]{\text{(integrated Sachs-Wolfe)}},
    \label{eq:Delta_g-FULL}
\end{alignat}
where \(\delta,\,\Theta,\,\kappa,\,\varPsi\), and \(\Phi\) are linear perturbations and \(b,\,\cH,\,\alpha\), and \(\bmag\) are redshift-dependent functions. Specifically, \(\delta\) is the matter density contrast (in the comoving-synchronous gauge), \(\Theta\) is the scalar potential of the peculiar velocity field \(\vec v\) (such that \(\vec v=\vec\nabla\Theta\)), \(\kappa\) the lensing convergence, \(\varPsi\) and \(\Phi\) are the two gauge-invariant Bardeen potentials, and \(\Upsilon\coloneqq(\varPsi+\Phi)/2\) is the Weyl potential, responsible for lensing. On the other hand, \(\cH\) is the conformal Hubble factor, \(b\) is the linear galaxy bias, \(\bmag\) is the magnification bias, \(\bevo\) is the evolution bias, and
\begin{equation}
    \alpha:=\frac{\cH'}{\cH^2}+2\,\bmag-2\,\frac{\bmag-1}{x_\|\,\cH}-\bevo
\end{equation}
is the amplitude of the Doppler term, itself crucially a function of \(\bmag\) and \(\bevo\). 
In \cref{eq:Delta_g-FULL}, we have neglected terms evaluated at $\vec x=0$ and those induced by the observer velocity 
\citep[see][for insights into the relevance of the proper motion of the observer]{2020IJMPD..2950085B,2021JCAP...11..027B,2022MNRAS.509.1626E}.

Several works have delved into the details of the derivation \citep[also including observer terms, as in][]{2018JCAP...10..024S,2020JCAP...11..064G}, and the significance of the corrections on the power spectrum \citep{2015CQGra..32d4001J,2020JCAP...09..058D,2020JCAP...07..048B,2022JCAP...01..061C,2023PhRvL.131k1201F}. 
\Cref{eq:Delta_g-FULL} displays a plethora of relativistic effects which have not been detected yet. In fact, they all act as subdominant contributions to the galaxy overdensity and only become relevant on the largest scales, as can be appreciated in Fourier space, where they explicitly manifest as either 
$\mathcal{O}(\cH/k)$ or $\mathcal{O}(\cH^2/k^2)$.
For this reason, efforts have been made to assess how relativistic corrections may affect a detection of the local Primordial non-Gaussianity parameter, $f_{\rm NL}$ \citep[see e.g.][]{2012PhRvD..85b3504J,2015MNRAS.451L..80C,2015ApJ...814..145A,2017PhRvD..96l3535A,2022MNRAS.510.1964M}, enlightening that a robust measurement of the latter should take into account the degeneracies with relativistic effects, at least those $\propto \cH/k$ \citep[for recent analyses, see e.g.][]{2021JCAP...12..004V,2021JCAP...11..010V,2024arXiv241206553G}. 

In the realistic case where the redshift bins have a non-null thickness, the angular power spectrum reads
\begin{equation} \label{eq:Cl_bin}
    C_\ell^{ij}=4\pi \int \frac{\de k}{k}\, \mathcal{P}(k) \,\Delta^i_\ell(k) \,\Delta^j_\ell(k)\,,
\end{equation}
where \(\mathcal{P}(k)\) is the dimensionless power spectrum of the primordial curvature perturbation, and
\begin{equation} \label{eq:Delta_ell}
    \Delta_\ell^i(k)= \int \de z\, n(z)\,W_i(z) \Delta_\ell(z,k)\,,
\end{equation}
where \(n(z)\) is the distribution in redshift of galaxies number counts per solid angle, \(i\) indicates the \(i\)th redshift bin, and \(W_i(z)\) is a window function, which can be of different shapes, one for each redshift bin. Assuming enough resolution in redshift measurements, we consider a tophat \(W_i(z)\). The last term in \cref{eq:Delta_ell} is given by
\begin{align}
    \Delta_{\ell}(z,k)=&\,j_{\ell}(kx_\|)\,b\,\delta(z,k) \nonumber \\ &\qquad\qquad\qquad\text{(density)}\nonumber \\
    &+j_{\ell}^{\prime \prime}(kx_\|)\,\frac{k}{\mathcal{H}}\,\Theta(z,k) \nonumber \\ &\qquad\qquad\qquad\text{(RSD)} \nonumber \\
    &+\left[j_{\ell}^\prime(kx_\|)\,\alpha
    +(\bevo-3)\,j_{\ell}(kx_\|)\,\frac{\mathcal{H}}{k}\right]\,\Theta(z,k) \nonumber \\ &\qquad\qquad\qquad\text{(Doppler)} \nonumber \\
    &+2\,(1-\bmag)\,\kappa_\ell(z,k) \nonumber \\ &\qquad\qquad\qquad\text{(lensing)}\nonumber \\
    &+j_{\ell}(kx_\|)\,\Phi'(z,k)/\mathcal{H} \nonumber \\ &\qquad\qquad\qquad\text{(Sachs-Wolfe)} \nonumber \\
    &+j_{\ell}(kx_\|)\left[(\alpha+1)\Psi(z,k) - 2\,(1-\bmag)\,\Phi(z,k)\right] \nonumber \\ &\qquad\qquad\qquad\text{(potentials)} \nonumber \\
    &+2\,(1-\bmag)\,\int^{x_\|\ }_0 \de r\,  j_\ell(kr)\, \frac{ \Upsilon(z,k)}{r(z)}
    \nonumber \\ &\qquad\qquad\qquad\text{(time delay)} \nonumber \\
    &+2\,\alpha\,\int^{x_\|\ }_0 \de r\,  j_\ell(kr)\,\Upsilon^\prime(z,k) \nonumber \\ &\qquad\qquad\qquad\text{(integrated Sachs-Wolfe).}
\end{align}
Above, the \(j_\ell\) are the spherical Bessel functions, primes denote derivatives w.r.t.\ either the argument of $j_\ell$ or conformal time (when applied to potentials), and
\begin{equation}
    \kappa_\ell(z,k)= \ell\,(\ell-1)\,\int^{x_\|\ }_0 \de r\,  j_\ell(kr)\, \frac{x_\|-r}{x_\|\,r}\,\Upsilon(z,k)\,
\end{equation}
is the convergence kernel in harmonic space.

\section{Multi-tracer analysis} \label{sec:MT}
From multiple observations of clustering in different redshift bins and two tracers of the underlying (dark) matter distribution, $t$ and $t'$, we can construct a multi-tracer tomographic matrix,
\begin{equation} \label{eq:MTtomographic_matrix}
    \tens C_\ell=\left(
    \begin{array}{cc}
        \tens C_\ell^{tt} & \tens C_\ell^{tt'} \\
        \tens C_\ell^{t't} & \tens C_\ell^{tt}
    \end{array}
    \right)\;,
\end{equation}
with \(\tens C_\ell^{tt}\) and \(\tens C_\ell^{t't'}\) respectively being the tomographic auto-spectrum matrix of the galaxy populations $t$ and $t'$, and \(\tens C_\ell^{tt'}\) their tomographic cross-spectrum matrix. Specifically, for the two samples subdivided into \(N_t\) and \(N_{t'}\) redshift bin, we have 
\begin{equation}
    \tens C_\ell^{tt'}=\left(
    \begin{array}{cccc}
        C_{11\ell}^{tt'} & C_{12\ell}^{tt'} & \ldots &
        C_{1N_{t'}\ell}^{tt'} \\
        \vdots & \vdots & \ddots & \vdots \\
        C_{N_t1\ell}^{tt'} & C_{N_t2\ell}^{tt'} & \ldots & C_{N_tN_{t'}\ell}^{tt'}
    \end{array}
    \right)\;,
\end{equation}
where \(C_{ij\ell}^{tt'}\) is the harmonic-space power spectrum between the tracer $t$ in the \(i\)th redshift bin, and the tracer $t'$ in the \(j\)th bin. Note that if the populations have a different number of redshift bins, the cross-spectrum matrix will be rectangular, whereas the auto-spectrum ones are always square.

Let now $t={\rm F}$ and $t'={\rm B}$ be, respectively, the faint and bright sub-samples of the same galaxy population. These (sub-)populations are independent once the luminosity cut is applied to the total catalogue; therefore, they can be seen as two independent realisations of the matter density field \citep{Bonvin_2014}. We can indeed apply the multi-tracer formalism outlined above to a single galaxy population, as the reader shall see in the remainder of this paper. 

To compute the theoretical expectation of any harmonic-space power spectrum, we employ the \texttt{CLASS} code \citep{2011arXiv1104.2932L,2011JCAP...07..034B}, which we modify in a number of ways to meet the requirements of our analysis. In the original \texttt{CLASS} version, the evolution bias \(\mathcal{E}\) was computed internally using a function called \texttt{n(z)\_evolution}; the same \(\mathcal{E}\) is now computed externally and passed to the code via an interpolation. In order to compute the multi-tracing tomographic matrix, we follow the method used in the Multi-CLASS code, implemented on the updated version of \texttt{CLASS} \citep{2020JCAP...10..016B,2020JCAP...10..017B}. 
This allows us to compute the square tomographic cross-spectrum matrix $C_\ell^{\rm FB}$ given a faint \(n(z)\) distribution, a bright \(n(z)\) distribution, their respective \(\bevo\), \(\bmag\) and \(b\), and a single redshift interval with its bin separation.
The \texttt{CLASS} code can use the Limber approximation \citep{1953ApJ...117..134L} to compute the $C_\ell$, starting from a selected angular multipole $\ell$. Although this method speeds up the computation, it could also produce numerical artifacts around the value of $\ell$ at which the code switches to the approximate method. In order to avoid these artifacts, we decide to use only the $C_\ell$ computed without the Limber approximation.

Our goal is to assess the statistical significance with which relativistic effects can be detected via a full multi-tracer analysis in harmonic space between a faint and a bright sub-sample. To this purpose, we split the model power spectrum into a `standard' term, containing the dominant contributions to the spectrum of \(\Delta\), and all the remaining components, namely
\begin{equation}
    \tens C_\ell=\tens C_\ell^{\rm(std)}+\tens C_\ell^{\rm(ext)}\;.
    \label{eq:model}
\end{equation}
Then, we construct the difference in chi-squared between an alternative hypothesis \(\chi^2_{\rm full}\), where we account for all relativistic contributions, and the null hypothesis \(\chi^2_{\rm std}\), in which we assume that the universe is described just by the dominant terms in \cref{eq:Delta_g-FULL}, i.e.\ there are no extra terms in \cref{eq:model} \citep[see also Sec.\ 4 in][]{2025A&A...697A..85E}.

For a synthetic, noiseless data set created by computing the multi-tracer tomographic power spectrum of \cref{eq:Delta_g-FULL}, we have, by construction, \(\chi^2_{\rm full}\equiv0\). Hence, \(\Delta\chi^2\equiv\chi^2_{\rm std}\), with
\begin{equation}
    \chi^2_{\rm std}=\sum_{\ell,\ell'=\lmin}^{\lmax}\left[\slashed{\vec C}_\ell^{\rm(ext)}\right]^{\sf T}\,\tens\Psi_{\ell\ell'}\,\slashed{\vec C}_{\ell'}^{\rm(ext)}\;,
    \label{eq:X2_vec}
\end{equation}
where \(\slashed{\vec S}\) denotes the half-vectorisation of symmetric matrix \(\tens S\), and \(\tens\Psi_{\ell\ell'}\) is the precision matrix for the data vector \(\smash{\hat{\slashed{\vec C}}_\ell}\), i.e.\ the matrix inverse of the covariance matrix \(\smash{\langle(\slashed{\vec C}_\ell-\hat{\slashed{\vec C}}_\ell)\,(\slashed{\vec C}_{\ell'}-\hat{\slashed{\vec C}}_{\ell'})^{\sf T}\rangle}\). In the Gaussian approximation, the latter is diagonal in \(\ell\)-\(\ell'\), and so is therefore the precision matrix.

After some lengthy matrix manipulation \citep[see e.g.][]{2008PhRvD..77j3013H}, it is possible to recast \cref{eq:X2_vec} as
\begin{equation}
    \chi^2_{\rm std}=\sum_{\ell=\lmin}^{\lmax}\tr\left[\tens C_\ell^{\rm(ext)}\,\tens\Sigma_\ell^{-1}\,\tens C_\ell^{\rm(ext)}\,\tens\Sigma_\ell^{-1}\right]\;,
\end{equation}
with \(\tens\Sigma_\ell\) representing the standard deviation on a measurement of \(\tens C_\ell\), namely
\begin{equation} \label{eq:covariance}
    \tens\Sigma_\ell=\sqrt{\frac2{(2\,\ell+1)\,\Delta\ell\,\fsky}}\,\left[\tens C_\ell+{\rm diag}(\vec{\bar n})^{-1}\right]\;,
\end{equation}
where \(\Delta\ell\) is the width of the multipole bins, \(\fsky\) is the fraction of surveyed sky, and \(\vec{\bar n}=\{\bar n_{{\rm F},i=1\ldots N_I}\}\cup\{\bar n_{{\rm B},j=1\ldots N_J}\}\) is the vector containing all average number densities of galaxies for each sub-sample and in each redshift bin, i.e.\ \(\bar n_{I,i}=\int\de z\,n_{I,i}(z)\) (in inverse steradians).

Having introduced all the relevant quantities, we proceed by computing the test statistics adopted as a function of the split between the faint and bright sub-samples. In previous works, this was done in terms of the splitting magnitude, \(m_{\rm s}\), or splitting flux, \(F_{\rm s}\), depending on survey specifications. Here, to avoid unnecessary complications, we opt instead to unify the treatment by introducing, as a proxy for the split, the faint fraction, \(f_{\rm F}\coloneqq \bar n_{\rm F}^{\rm(tot)}/\bar n_{\rm T}^{\rm(tot)}\), with \(\smash{\bar n_I^{\rm(tot)}=\int_{z_{\rm min}}^{z_{\rm max}}n_I \,\de z}\).
Then, we present results in terms of \(\Delta\chi^2(f_{\rm F})\), so that it will be clear whether and when an asymmetric division is preferred \citep[this choice will also ease comparison with other works, e.g.][]{2023arXiv230604213B}.

To assess the relevance of $\tens C_\ell^{\rm(ext)}$, we pick three galaxy targets at different redshifts. Given the intrinsic sample dependence of the power spectrum, luminosity functions must be used to estimate the number density and biases of each population. At low-$z$, we focus on a DESI-like BGS \citep{2024AJ....167...62D,2023AJ....165..253H}, which we model by implementing the Halo Occupation Distribution fits presented in \citet[][as it has been done in Appendix A.2 of \citealt{2023arXiv230912400M}]{2023arXiv231208792S}. At intermediate redshifts, we consider a population of H$\alpha$ emitters, adopting the \textit{Model 3} luminosity function in \citet[][see also \citealt{2021JCAP...12..009M}]{2016A&A...590A...3P}, which has proven to be a good description of the main spectroscopic target of the \textit{Euclid} satellite \citep{2024arXiv240513491E}. Finally, we model an LBG sample \citep[][see also Appendix D in \citealt{2024arXiv240706301R}]{2019JCAP...10..015W,2021JCAP...12..049S}, in order to mimic the planned MegaMapper \citep{2019BAAS...51g.229S,2022arXiv220904322S} survey.

In order to assess the intrinsic presence of observable GR effects in each galaxy population, we shall assume $f_{\rm sky}=1$. The reader will be able to rescale our result thanks to \cref{eq:covariance}, using $f_{\rm sky}\simeq0.36$ for DESI- and \textit{Euclid}-like surveys, and $f_{\rm sky}\simeq0.5$ for a MegaMapper-like experiment. This permits avoiding results driven by survey's sky coverages, ensuring our forecasts for different samples to be immediate to compare. 
Crucially, biases for the faint sub-sample are affected by both the lower and the upper luminosity thresholds. We therefore model them in line with \citet{2023arXiv230604213B} and \citet{2023arXiv230912400M}.

\section{Results}\label{sec:results}
The vast majority of cosmological analyses employing galaxy clustering statistics \citep[e.g.][]{2025JCAP...07..028A} focusses on the first two terms in \cref{eq:Delta_g-FULL}, i.e.\ matter density fluctuations and RSD. Lensing magnification is customarily not included, for two main reasons, both related to it being an integrated effect. First, lensing becomes important at high redshift, and most galaxy redshift surveys have hitherto focused on measurements at \(z\lesssim1\). Secondly, integrated terms are inherently difficult to model in Fourier-space analyses, such are those performed when spectroscopic redshifts are available \citep[see e.g.][]{2016JCAP...01..016D}, and the `non-locality' of the lensing contribution makes harder to understand to which wavenumbers in the power spectrum it mostly contributes.

The situation is different when dealing with summary statistics in polar coordinates, such as the angular two-point correlation function and the harmonic-space power spectrum. In this case, both local and integrated terms are well-defined and well-behaved, and can be included accordingly \citep[see e.g.][]{2023MNRAS.523.3649E,2021A&A...646A.140H,2022A&A...662A..93E}. For this reason, we shall always include in \(\smash{\tens C_\ell^{\rm(std)}}\) density, RSD, and magnification.

On the other hand, we shall consider two distinct cases regarding relativistic effects, depending on whether the Doppler term is included in \(\smash{\tens C_\ell^{\rm(std)}}\) or rather in \(\smash{\tens C_\ell^{\rm(ext)}}\). The reason for the latter is obvious, whereas the reason for the former lies in the fact that the Doppler term is the strongest of all relativistic contributions to \(\Delta\), thanks to its scaling with \(\cH/k\). As such, its measurement, albeit not attained yet, is foreseen in the very next years, thanks to the oncoming data from surveys such as DESI and \textit{Euclid}. All other relativistic contributions (fourth to seventh rows in Eq.\ \ref{eq:Delta_g-FULL}) scale instead with \(\cH^2/k^2\), which makes them largely subdominant, even with respect to Doppler. For this reason, we shall henceforth either refer to a detection of the Doppler term, if it is included in \(\smash{\tens C_\ell^{\rm(ext)}}\), or to that of 2nd-order (in $\smash{\cH/k}$) relativistic corrections, if it is included in \(\smash{\tens C_\ell^{\rm(std)}}\).

\begin{figure}
    \centering
    \includegraphics[width=\columnwidth]{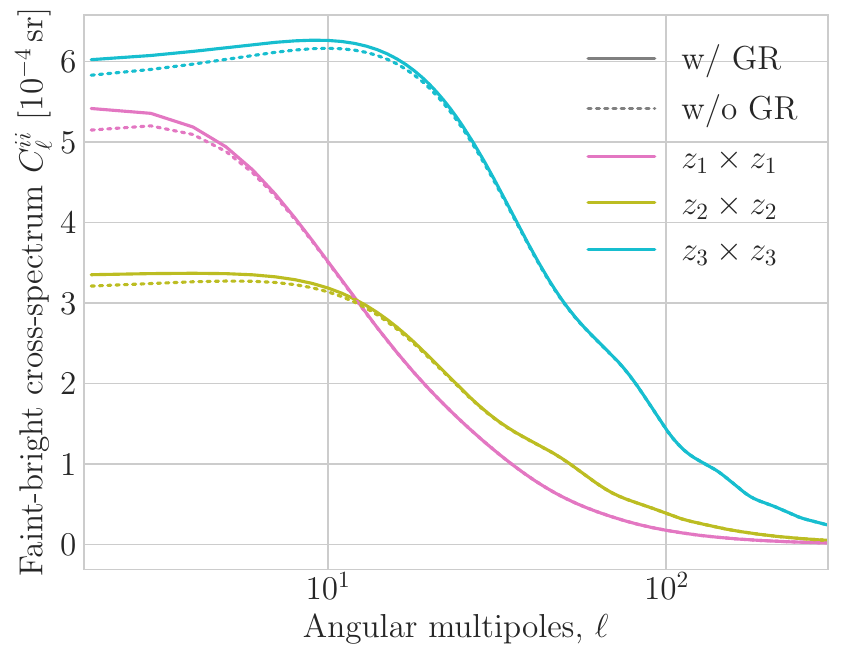}
    \caption{Harmonic-space power spectrum for faint-bright cross-correlation in the case of BGS. All spectra are plotted in the range $\ell \in [2,300]$. Solid curves correspond to the inclusion of all the contributions, whereas dashed ones correspond lack Doppler and the other GR terms. By $z_i$ we indicate the $i$th redshift bin, so that the $z_i \times z_i$ curve is the auto-correlation power spectrum in the $i$th bin, i.e.\ $C^{ii}_\ell$.}
    \label{fig:Cl}
\end{figure}

\begin{figure*}
    \centering
    \includegraphics[width=\textwidth]{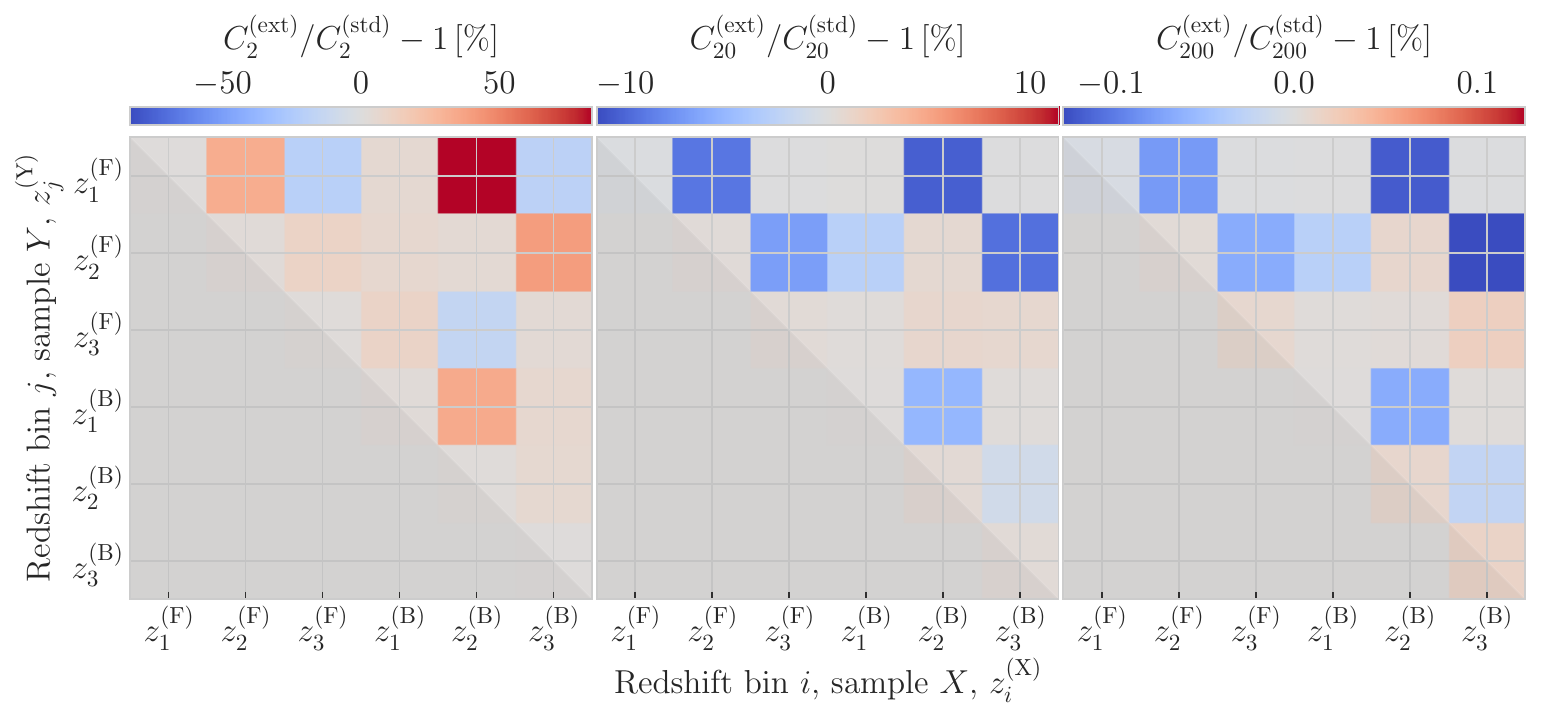}
    \caption{Percentage change between the multi-tracer faint-bright tomographic matrix that includes all GR effects (i.e.\ the Doppler term is taken to be part of ${\sf C_\ell^{\rm (ext)}}$ here) and the one that does not. Three fixed multipoles are considered: $\ell=\{2,\,20,\,200\}$, from left to right. In line with \cref{fig:Cl}, we use the BGS specs, with $3$ redshift bins.}
    \label{fig:tomographic_m}
\end{figure*}

For illustrative purposes, we show in \cref{fig:Cl} the angular power spectrum of the cross-correlation between the two sub-samples. We limit to the auto-correlation of a $3$ redshift bins configuration of our DESI-like BGS, which we display both including (solid lines) and not including (dashed lines) relativistic corrections. The reader can easily notice the differences between ``w/\ GR" and ``w/o GR", which only acquire significance at low-$\ell$.
To further show how relativistic effects impact on angular power spectra, \cref{fig:tomographic_m} depicts the percentage variation they induce on the full tomographic matrix, in \cref{eq:MTtomographic_matrix}, at: ultra-large scales ($\ell=2$, left panel), intermediate scales ($\ell=20$, central panel), and small scales ($\ell=200$, right panel). Once again, we stress the importance of looking at the lowest multipoles when seeking a relativistic signature. Also, it is worth noticing that GR affects the cross-correlations of neighbouring redshift bins the most, and that, generally speaking, the cross-correlation between the two sub-samples is more influenced than the corresponding auto-correlations. 

\subsection{Optimisation w.r.t.\ faint/bright split}
\begin{figure}
    \centering
    \includegraphics[width=\columnwidth]{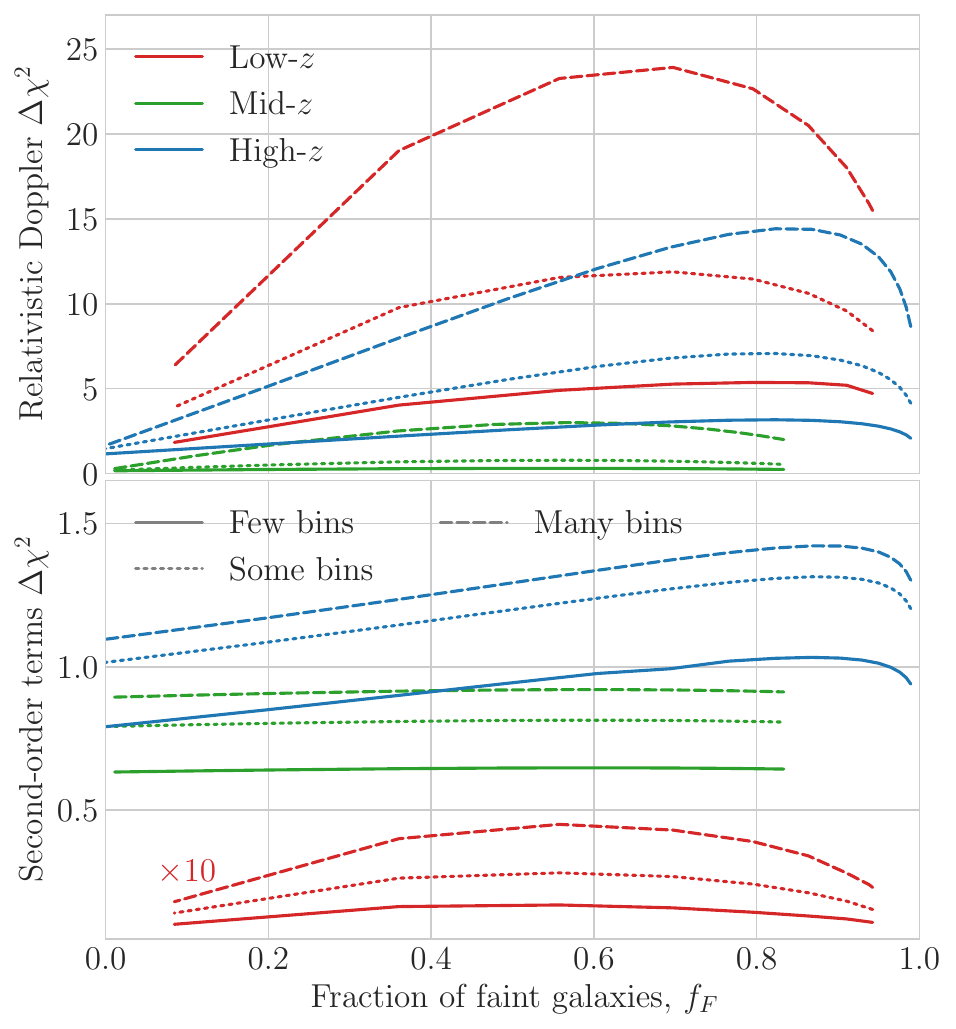}
    \caption{Cumulative statistical significance associated with a detection of the relativistic contributions as a function of the fraction of faint galaxies. Low-$z$ DESI-like BGS curves in red, intermediate-$z$ \textit{Euclid}-like H$\alpha$ in green, and high-$z$ MegaMapper LBGs in blue. Different binning choices are shown via the line-style code, increasing the number of redshift bins going from the solid lines, through dotted lines, to dashed lines. \textit{Top:} $\Delta\chi^2$ for a model with all the GR effects against a null-hypothesis of a `standard' angular power spectrum, i.e.\ Doppler included in $\smash{{\sf C}_\ell^{\rm (ext)}}$. \textit{Bottom:} $\Delta\chi^2$ of the GR corrections $\mathcal{O}(\cH^2/k^2)$, i.e.\ Doppler in $\smash{{\sf C}_\ell^{\rm (std)}}$. The statistical significance of BGS is enhanced by a factor of $10$ to enhance readability.} 
    \label{fig:X2}
\end{figure}

Splitting a galaxy catalogue into a faint and a bright sub-samples, first proposed by \citet{Bonvin_2014} \citep[see also][]{2014PhRvD..89h3535B,2017JCAP...01..032G,2023arXiv230604213B,2023arXiv230912400M,2024PDU....4601634M}, offers a way to boost the signal due to the subdominant contribution from relativistic effects. 
In this article, for the first time in harmonic space, we forecast the cumulative---namely, over the entire redshift range---statistical significance $\Delta\chi^2(f_{\rm F})$ of both the Doppler (\cref{fig:X2}, upper panel) and the 2nd-order relativistic (\cref{fig:X2}, lower panel) terms. 

We plot $\Delta\chi^2(f_{\rm F})$ for three binning scenarios---a wide (solid lines), an intermediate (dashed lines) and a narrow (dotted lines) bin case---for the galaxy populations under consideration, in \cref{fig:X2}. Results for BGS are shown in red (and incremented by a factor of $10$ to enhance readability in the lower panel), for H$\alpha$ ELGs in blue, whilst green is used for LBGs. 
The Doppler term is overall dominant and exhibits $\Delta\chi^2$ values of roughly an order of magnitude higher than those of second-order contributions. It also proves to be of the utmost significance at low-$z$, due to the factor $\propto (r\,\cH)^{-1}$ present in the Doppler amplitude $\alpha$. Moving to higher redshifts, we observe that the features induced by the galaxy populations become more relevant than the redshift dependence, with the LBG sample being preferred over H$\alpha$ ELGs because of its higher biases and larger accessible volume. 

Whilst the significance of the Doppler term is greatly affected by the number of redshift bins considered, our results for the second-order corrections are mostly affected by the total redshift interval.
Indeed, such contributions require huge survey volumes, as well as high target number densities, to be measured. Also, the higher the redshift, the longer the line of sight along which we integrate in the case of time-delay and ISW effects. 
We find $\Delta\chi^2\approx1$ in the case of LBGs, $\Delta\chi^2\approx0.5$ for H$\alpha$, and  $\Delta\chi^2\approx0.03$ for BGS.
Therefore, we cannot forecast a detection of the second-order contributions for any of the galaxy samples under consideration. 
On top of that, in both panels of \cref{fig:X2}, we can appreciate the actual dependence of the statistical significance upon $f_{\rm F}$, which enforces the idea \citep[already present in the literature, see e.g.][]{2023arXiv230604213B,2023arXiv230912400M,2025PhRvD.111l3559C} that a tailored luminosity cut may increase the probability of detecting the GR contributions. In particular, we find the following optimal configurations: $f_{\rm F}\approx0.6$, $0.5$ and $0.9$ at low-, mid-, and high-$z$, respectively. 
\review{It may be difficult to appreciate the dependence on $f_{\rm F}$ in the mid-$z$ curves in \cref{fig:X2}. In the top panel, this difficulty is only due to the scale of the plot, whereas in the bottom one the dependence is, in fact, less prominent. The reason for this weak dependence is due to the small impact that $\bmag$ and $\bevo$ have on the subdominant GR effects in that configuration.}

Since the harmonic-space power spectra project clustering properties on bidimensional spherical shells, information along the line of sight direction can only be retrieved by taking finer redshift bins \citep[see also][]{2013JCAP...11..044D,2018MNRAS.481.1251C}. Therefore, if we go from a thick- to a thin-bin scenario, we will increase the signal-to-noise ratio of the relativistic effects, as can be noticed in \cref{fig:X2} for all targets. 
In particular, we move from $3$ redshift bins (solid lines) to $5$ (dotted lines) and then $7$ (dashed lines) bins in the case of BGS. Similarly, we consider $4$, $8$ and $32$ for H$\alpha$ ELGs, and $7$, $14$ and $28$ bins for LBGs.

Finally, we remind the reader that \cref{fig:X2} refers to the idealistic case of $f_{\rm sky}=1$. Using $\Delta\chi^2\propto f_{\rm sky}$---as \cref{eq:covariance} shows---for realistic sky coverages, we find $\Delta\chi^2\lesssim8,\,1.5$ and $7$, respectively for a DESI-like BGS, an \textit{Euclid}/-like H$\alpha$ target and a MegaMapper-like LBGs sample, when relativistic Doppler is considered as part of \(\smash{\tens C_\ell^{\rm(ext)}}\). On the other hand, when Doppler \(\smash{\subset\tens C_\ell^{\rm(std)}}\), we have $\Delta\chi^2\lesssim0.02$ (BGS), $\Delta\chi^2\lesssim0.3$ (H$\alpha$ ELGs), and $\Delta\chi^2\lesssim0.7$ (LBGs). 

\review{It is important to notice that a simpler study made without splitting the galaxy catalogues into two sub-samples would yield the same results as in \cref{fig:X2} in the limit $f_{\rm F}=0$. In such a configuration, far from the optimal one, the bright galaxy sample would contain all the galaxies in the catalogue, while the faint galaxy sample would be completely empty.}

\subsection{Sub-dominant relativistic contributions}
Our results clearly show that the main contribution to the statistical significance comes from the Doppler, which is likely to be detected with the proposed strategy. On the other hand, the second-order GR contributions only attain relevance at high-$z$, and are safely negligible otherwise. Yet a $\Delta\chi^2\simeq1$ does not suffice to claim a possible detection with the upcoming surveys. Two possible conclusions are suggested by this outcome: either second-order terms need even denser galaxy catalogues, or the faint-bright multi-tracer technique itself prevents them from getting high relevance, due to the relationship between the biases of the sub-samples. 

We then choose to assess their statistical power in such measurements, since our modelling framework enables us to work out the average galaxy density required to have at least a $3\,\sigma$ detection. 
To this purpose, we adopt an agnostic $50\%-50\%$ faint-bright MegaMapper configuration with thin redshift bins, and let the shot-noise term that enters the covariance vary, keeping for simplicity all of the other sample-dependent quantities fixed. This corresponds to a quasi-optimal choice since, whilst it refers to the best-performing binning scenario, it does not coincide with the peak of the $\Delta\chi^2(f_{F})$ curve in \cref{fig:X2}.
Such an exercise shows, as expected, that the signal increases when reducing the noise level; however, $\Delta\chi^2\simeq9$ would correspond to $\bar n^{\rm (tot)}\gtrsim 2\,000 \; \rm arcmin^{-2}$, which is a totally unrealistic value.
One could refine this estimate by considering the variation in $b$, $\bmag$ and $\bevo$ induced by the lower critical flux of a survey displaying a much higher galaxy number density. However, we point out that such a calculation would require the extrapolation of the luminosity function model down to regimes where it has not been calibrated, and a characterisation of target samples for futuristic surveys is beyond the scope of our work.

As a consequence, we conclude that also in the future $\mathcal{O}(\cH^2/k^2)$ effects will make up only a small fraction of the relativistic signal within faint-bright angular power spectrum analyses. This does not come as completely unexpected, since it is in line with what previous works have found, even looking at combinations of surveys \citep{2021JCAP...11..010V}. To look at them, probably alternative approaches will be needed, involving different datasets with larger bias differences \citep[see e.g.][]{2025arXiv250515645Z}. Nevertheless, we can put this result into another perspective and argue that a real measurement using the luminosity cut may be safely done by focusing on a simpler model that only includes the Doppler correction.

\section{Conclusions}\label{sec:conclusions}

A measurement of GR effects on cosmological scales would be a great confirmation of the validity of Einstein's theory in a regime where it is currently poorly tested. Current and upcoming galaxy surveys could observe these effects for the first time, thanks to the size of the probed volume and the number of galaxies contained in it. In this work, we have presented forecasts for the detection of relativistic effects with angular power spectrum measurements, using internal cross-correlation of various galaxy samples. Following \citet{2014PhRvD..89h3535B,2023arXiv230604213B,2024PDU....4601634M}, we have defined sub-samples to be cross-correlated employing a luminosity cut. Since the amplitudes of these effects vary depending on the considered galaxy population, we have optimised such samples in order to find the best configuration for the detection of relativistic corrections. In particular, we have analysed three tracers of the cosmic large-scale structure: a sample of bright galaxies at low redshift (BGS, DESI-like), a sample of emission-line galaxies at medium redshift (H$\alpha$, \textit{Euclid}-like) and a LBG sample at high redshift (MegaMapper-like). 

With the BGS sample, by carefully selecting the fraction of faint galaxies $f_{\rm F}$, we have obtained a $\sim3\,\sigma$ Doppler detection significance with a maximum at about $f_{\rm F}=0.7$, even in a realistic case of $f_{\rm sky}\simeq0.36$. A similar result has been obtained using the LBG sample, with a maximum at about $f_{\rm F}=0.8$. If in the BGS case the high significance is due to the fact that Doppler contribution is dominant at low redshift, in the LBG case, the high significance is due to the size of the probed volume and the high number of redshift bins considered. Looking at the high detection significance we have forecasted, we can state that a bright galaxy survey (DESI-like) is the preferred target to consider. At the same time, we have found the Doppler significance in the H$\alpha$ emitters case to be much lower than the $3\,\sigma$ target, making \textit{Euclid}-like experiments not ideal for observing relativistic Doppler.

Focusing on the detection significance of the sub-dominant relativistic effects, we have not been able to find a test configuration in which it reached the $3\,\sigma$ target. 
Nevertheless, our findings show that the detection significance has an interesting behaviour depending on the kind of survey we have considered: it does not increase as much as in the relativistic Doppler study when the number of redshift bins increases, but it does improve with the total probed volume, i.e.\ the total redshift interval of the survey. To see if a larger experiment would be able to measure the $\mathcal{O}(\cH^2/k^2)$ relativistic effects, again via the faint-bright split strategy, we have calculated, with a rough estimate, the average number of galaxies needed to reach the $3\,\sigma$ level. However, this estimated has led to $\bar n^{\rm (tot)}\gtrsim 2\,000 \; \rm arcmin^{-2}$. Since this value is far from being realistic, we conclude that, in order to have a detection of sub-dominant contributions, a different approach should be applied.

In conclusion, the use of a power spectrum analysis in harmonic space confirms that the Doppler effect is, in fact, the dominant correction to the galaxy number density. Other $\mathcal{O}(\cH^2/k^2)$ terms are not detectable with the presented method. In future studies, the multi-tracing technique could be expanded by splitting the single galaxy population into three or more sub-samples. While, number of galaxies in each sample will certainly decrease, thus increasing the noise, it will be interesting to seek additional information about the sub-dominant relativistic effects within the additional cross-correlations.


\section*{Acknowledgments}
The authors acknowledge support from the Italian Ministry of University and Research (\textsc{mur}), PRIN 2022 `EXSKALIBUR – Euclid-Cross-SKA: Likelihood Inference Building for Universe's Research', Grant No.\ 20222BBYB9, CUP D53D2300252 0006, from the Italian Ministry of Foreign Affairs and International Cooperation (\textsc{maeci}), Grant No.\ ZA23GR03, and from the European Union -- Next Generation EU.

\bibliographystyle{apsrev4-1}
\bibliography{apssamp}


\end{document}